\documentclass[twocolumn,aps,prb,floatfix,showpacs]{revtex4}
\usepackage{graphicx,bm}

\begin{document}

\title{
Ferromagnetism in graphene traced to an antisymmetric orbital combination of 
involved electronic states
}

\author{Wei Xu}
\author{J. G. Che}\altaffiliation{Corresponding author.
E-mail: jgche@fudan.edu.cn.}
\affiliation{Surface Physics Laboratory (National Key Laboratory),
Key Laboratory of Computational Physical Sciences (MOE),
Department of Physics and
Collaborative Innovation Center of Advanced Microstructures, Fudan University,
Shanghai 200433, People's Republic of China}


\begin{abstract}
Based on first principles calculations, we reveal that the origin of 
ferromagnetism caused by $sp$ electrons in graphene with vacancies 
can be traced to electrons partially filling $sp^{2*}$-antibonding and 
$p_z^*$-nonbonding states, which are induced by the vacancies and appear 
near the Fermi level. Because the spatial wavefunctions of the both states 
are composed of atomic orbitals in an antisymmetric configuration, their 
spin wavefunctions should be symmetric according to the electron exchange 
antisymmetric principle, leading to electrons partially filling these states 
in spin polarization. Since this $p_z^*$ state originates not from 
interactions between the atoms but from the unpaired $p_z$ orbitals 
due to the removal of $p_z$ orbitals on the minority sublattice, 
the $p_z^*$ state is constrained, distributed on the atoms of 
the majority sublattice, and decays gradually from the vacancy 
as $\sim$ $1/r$. According to these characteristics, we concluded that 
the $p_z^*$ state plays a critical role in magnetic ordering in graphene 
with vacancies. If the vacancy concentration in graphene is large enough 
to cause the decay-length regions to overlap, constraining the $p_z^*$ 
orbital components as little as possible on the minority sublattice atoms 
in the overlap regions results in the vacancy-induced $p_z^*$ states being 
coherent. The coherent process in the overlap region leads to 
the wavefunctions in all the involved regions antisymmetrized, 
consequently causing ferromagnetism according to the electron exchange 
antisymmetric principle. This unusual mechanism concerned with 
the origin of $sp$-electron magnetism and magnetic ordering 
has never before been reported and is distinctly different 
from conventional mechanisms. Consequently, we can explain 
how such a weak magnetization with such a high critical temperature 
can be experimentally observed in proton-irradiated graphene.
\end{abstract}

\maketitle

\section{Introduction}
\label{Introduction}
The present work examines the origin of magnetism in nominally nonmagnetic materials with 
only $sp$-electrons, which is the one of the most controversial issues in modern materials 
science\cite{Mak06,Yaz10,Esq13,Fen17,Naf17}, by considering graphene containing vacancies as 
an example. Magnetism induced by removing a single $p_z$ orbital from the $\pi$-electron 
systems, such as graphene or graphite, have been observed by experiments at room temperature 
and have been predicted by calculations.\cite{Kuz13,Esq13,Han14,Fis15}. 
However, it is still quite controversial. Nair {\it et al.} reported 
that no magnetic ordering could be detected down to liquid helium temperatures\cite{Nai12}.
This phenomenon has 
attracted a significant amount of attention owing to its potential applications in spintronics, 
nanostructures and biocompatible materials. Before they can be regarded 
as candidate materials for the applications, a 
comprehensive understanding of the origin and coupling of the magnetic moments (MM) in these 
materials is required\cite{Mak06,Fen17}. However, the basic theory concerning magnetism in 
solids established by Heisenberg in the 1920's\cite{Hei28} stressed that to cause magnetism 
in solids, the principal quantum number of the electrons must be greater than or equal to three 
($n\ge 3$). Therefore, 40 years ago, when observations of magnetism in light-element materials 
(containing only $sp$ electrons, $n < 3$) were reported, the first response was that these samples 
might be contaminated by magnetic impurities\cite{Gar06}. In recent years, with the reports of 
observed magnetism in such materials having continuously increased and carefully analyses 
having excluded impurities as their magnetic origin~\cite{Esq02,Mak06,Uge10,Esq13}, the 
evidence for ferromagnetic properties in these nominally nonmagnetic materials is considered firm. 
However, the magnetic origin or the mechanism responsible for the magnetism in these nominally 
nonmagnetic materials remains rather 
unclear~\cite{Coe02,Mak06,Yaz10,Kat12,Kuz13,Sin13,Han14,Fis15,Fen17,Naf17}. 

Even though it was still debated for FM or PM in irradiated graphene, 
we would focus on the serious difficulties met by theoretical study 
in the following two fronts: first, the origin of 
magnetism in $sp$ electron materials ($n < 3$), and second, how such small magnetization (three 
or four orders of magnitude smaller than conventional magnets) and such long-range magnetic 
coupling (a distance of 20~\AA~ between vacancies) can be ferromagnetic at room 
temperature\cite{Esq03,Cer09,Wan09,Fis15}. 

On the first front, the current explanation is based on Hund's 
rule~\cite{Leh03,Leh04,Per06,Yaz07,Pal08,Lis10,Pal12,Nan12} because the states to which 
magnetism can be traced are induced by vacancies, which help localize $sp$ electrons, such as 
those in isolated atoms. If this magnetism is observed in experiments, it could be understood using 
Hund's rule because atom-like-localized electrons could actually exist in materials. However, if 
this magnetism emerges from calculations based on band theory and on the singlet-electron 
approximation (SEA), one needs to be careful with the interpretation. Within the framework of 
band theory, the eigenstates are Bloch modes that are all extended according to the Bloch 
theorem~\cite{Bloch}. As such, the magnetism arising from calculations based on band theory 
cannot be attributed to the localized nature of electrons per se because localized electrons do not 
exist in calculations based on SEA and the Bloch theorem. Even if the dispersion of a band looks 
flat, the electrons on this band are still extended, moving everywhere in crystal~\cite{Bloch}. 

On the second front, if Heisenberg's model is used to explain magnetic ordering in such materials, 
the contradictions exist for weak magnetization (three or four orders of magnitude smaller than 
that in classic magnets such iron), long-range magnetic orders (that is, a low vacancy 
concentration\cite{Elf02,Uge10,Esq13}), and high critical temperature (higher than room 
temperature)\cite{Esq03,Cer09,Wan09,Esq13}. Note that other models for this subject, such as the 
superexchange model, the double exchange model and the RKKY model,
can essentially be seen as extensions of the Heisenberg model, 
because they all have a term of a scalar product of total spin moments on atoms $i$ 
and $j$, $\mathbf{S}_i\cdot\mathbf{S}_j$. 
The differences between these models lie only in the coupling method. 

Clearly, there must be more substantial physics behind the magnetism in nominally nonmagnetic 
materials (without $d$ or $f$ electrons) and there must be an unusual mechanism different from the
conventional mechanism in $df$-electron solids.
This is a great challenge and requires a root-and-branch rework of magnetic theory\cite{Kuz13}. 

In the present study, we report that the magnetism in graphene with vacancies can be traced to 
electrons partially occupying $sp^{2*}$-antibonding states or $p_z^*$-nonbonding states because 
the spatial wavefunctions of the both states are antisymmetric and their spin wavefunctions should 
be symmetric according to the electron exchange antisymmetric principle. This is the origin of the 
magnetism induced by $sp$ electrons. It has been recognized that the nonbonding state (or zero 
mode in literature\cite{Per06,Per08}) decays with $\sim$ $1/r$ ($r$ being the distance to the vacancy) 
with long-range interaction\cite{Per06,Per08}. However, it is still unclear how the long-range 
interaction between these states created by vacancies can be so strong to couple such weak 
magnetic moments and reach such a high transition temperature (higher than room 
temperature)\cite{Esq03,Cer09,Wan09,Esq13}. This is obviously in contrast to the conventional 
magnetic theory based on Heisenberg's exchange model\cite{Hei28}. 
Induced not by interaction but by unpaired $p_z$-electrons, the 
nonbonding states themselves cannot strongly interact with each other. Regarding the long-range 
and antisymmetric natures of the nonbonding state, 
combined with 
the experimental and theoretical observations\cite{Kop00,Coe02,Per08,Pis08,Cer09,Wan09,Sep10,Nan12,Gon16},
we propose that the 
nonbonding state plays an important role in the ferromagnetism observed in graphene with 
vacancies. Due to its unpaired nature, the wavefunction of the state extends long range and keeps 
the orbital components on the atoms of the minority sublattice as little as possible. If the vacancy 
concentration in the minority sublattice of graphene is sufficiently large that the distances between 
the vacancies are smaller than the decay-length of the nonbonding states, the 
nonbonding-state-involved regions can be overlap. The induced nonbonding states could then be 
coherent in order to keep their orbital components on the atoms of minority sublattice as little as 
possible. Nonbonding states coupled in this way would maintain antisymmetric spatial 
wavefunctions, leading to ferromagnetic ordering.

These states (labeled by $sp^{2*}$ and $p_z^*$ in the present work) contributing to magentism were 
also reported in the previous investigations~\cite{Ma04,Per06,Yaz07,Per08,Pis08,Nan12,Pal12}. 
In these papers, however, the magnetism of these states was attributed to the localization 
(for the unsaturated dangling bond states due to the so-called Jahn-Teller distortion (JTD)) according to the Hund's rule 
and to the itinerant electron (for the zero-mode states due to unpairing pi-electrons between two sublattices) 
according to the itinerant electron mode, see e.g. the review article \cite{Yaz10}. 
In contrast, the present work, for the first time, traced the magnetism in nonmagnetic materials to 
the antisymmetric manner of the spatial wavefunctions of the involved electronic states according to 
the electron exchange antisymmetric principle of quantum mechanics: 
if a spatial wavefunction of an electronic state is antisymmetric, 
its spin wavefunction must be symmetric. 
We believe this undebatable principle can end a long history of dispute about the origin of magnetism 
in nonmagnetic materials.

\section{Calculation methods} 
\label{methods}
The analyses were performed using our first principles calculations concerning the frame work of 
the spin density functional theory as implemented in the Vienna Ab initio Simulation Package 
(VASP)~\cite{vasp}. Electron-ion interactions were described using the projector augmented 
plane wave method~\cite{PAW}. It is well established that the generalized gradient approximation 
(GGA) of exchange-correlation functional is favorable for treating systems with non-uniform 
charge densities such as graphene. Therefore, the Perdew-Burke-Ernzerhof form of 
GGA~\cite{GGA-PBE} was adopted in our calculations. In addition, we tested several cases using 
the local spin density functional (LSDA) for exchange-correlation effects~\cite{LDA}, which did 
not change the conclusions obtained with GGA. The wave functions were expanded in a plane 
wave basis with an energy cutoff of 500~eV throughout the calculations. Two-dimensional 
Brillouin-zone (BZ) integrals were sampled on k-meshes corresponding to 48$\times$48 and 
96$\times$96 in a 1$\times$1-sized graphene cell when calculating the total energy and the 
density of states (DOS), respectively. The equilibrium lattice constant obtained via total energy 
minimization was 2.468\AA~ for graphene, in good agreement with the experimental value, 
2.46~\AA~\cite{Net09}. The systems were modeled as supercells with a vacuum of 
approximately 20~\AA. All atoms in the slab were allowed to relax until the Hellmann-Feynman 
forces on the atoms were smaller than 0.01~eV/\AA. This calculation setup was found to be 
sufficiently accurate for our study. The maximally localized Wannier function process 
implemented in Wannier90 package\cite{Wan90} was performed to obtain the orbital information 
for the relevant electronic structures. 

\section{Results and discussion}
\label{results}
\subsection{A short summary and statements for the present study} 
\label{summary}
The main results (the MM and the long and short distance between three atoms surrounding the 
vacancy in the ground state) of graphene with a single vacancy in different vacancy concentration 
(for 2$\times$2$\sim$8$\times$8-sized unit cells) are summarized in Table I. The MM for the cases 
vary over a range of 1.0 and 1.6 Bohr magneton($\mu_{\rm B}$) and reach a stable value in the 
7$\times$7-sized cell, in good agreement with the corresponding results of previous 
calculations~\cite{Per06,Per08,Nan12}.

\begin{table}[h]
\caption{
Magnetic moments (MM) and structural parameters SD (short distance) and LD (long 
distance) describing the Jahn-Teller distortion for graphene with a vacancy in 2$\times$2 $\sim$
8$\times$8-sized cell.
}
\label{sumary}
\begin{ruledtabular}
\begin{tabular}{cccccccc}
Size cell & 2$\times$2 & 3$\times$3 & 4$\times$4 & 5$\times$5 & 6$\times$6 & 7$\times$7 & 8$\times$8 \\
\hline
MM $(\mu_B)$ & 1.0 & 1.1 & 1.6 & 1.6 & 1.6 & 1.3 & 1.3\\
SD (\AA) & 2.59 & 2.56 & 2.18 & 2.11 & 2.05 & 1.99 & 1.97\\
LD (\AA) & 2.59 & 2.56 & 2.58 & 2.58 & 2.58 & 2.57 & 2.57 
\end{tabular}
\end{ruledtabular}
\end{table}

In Table I, the JTD can be identified by the long and short sides 
of the isosceles triangle formed by the three atoms surrounding the vacancy. If the lengths of the 
long and short sides are equal, there is no JTD. Cases with cell sizes of 2$\times$2 and 3$\times$3 
do not have JTD. To check whether JTD occurs in these cases, the symmetry of $C_{\rm 3V}$ was 
destroyed by an artificial distortion (the length of the short and long sides being 1.40~\AA~ and 
2.25~\AA, respectively) as the initial atomic structure in the calculations. However, the optimized 
structures for the cases of 2$\times$2 and 3$\times$3 show that the differences between the short 
and long side disappear within our calculation accuracy, indicating that these two cases 
energetically favor keeping the $C_{\rm 3V}$ symmetry without JTD. It can be understood because within a 
small sized cell, the obtained energy due to JTD cannot compensate for the energy lost due to 
strain. The magnetic moments exist for all the investigated cases, varying between 1.0~$\mu_{\rm 
B}$ and 1.6~$\mu_{\rm B}$, indicating that the magnetism in graphene with vacancies does not 
depend on the existence of JTD. 

Even though the vacancy concentrations in which magnetism in graphene can be experimentally 
observed~\cite{Elf02,Uge10} were much smaller than our investigated 2$\times$2 and 
3$\times$3-sized cells, we would still take them into account to reveal the origin of the magnetism, 
because there is a structural turning point (with or without JTD). The JTD essentially does not 
affect the $p_z$ states but does significantly affect the $sp^2$ states relative to interactions 
between dangling bonds. The JTD-induced $sp^{2}$ dangling bond state has been understood in 
previous studies~\cite{Ma04,Yaz07,Nan12,Sin13} as an origin of magnetism in graphene 
with vacancies, that is, the $sp^{2}$ dangling bond state, which is induced by JTD in graphene 
with vacancies, contributed 1~$\mu_{\rm B}$ according to Hund's rule because the dangling bond 
state looks like the state of an isolated atom. However, we will show later that, even if no JTD 
exists for the 2$\times$2 and 3$\times$3 sized cases, the $sp^{2}$ states of the three atoms 
surrounding the vacancy also contribute to the magnetism with quite different mechanism from 
that in $df$-electron magnets. From the evolution of the MM relative to the electronic states near 
the turning point, the origin of the magnetism in the materials with only $sp$-electrons can be 
unambiguously understood, as discussed in more detail later. Therefore, the cases of the $2\times 
2$ and 3$\times$3 sized cells are also considered in this study. 

Note that all the atomic configurations listed in Table I are planar. Even though it has also been 
reported from first principles calculations~\cite{Pad16} that a nonplanar metastable state exists, 
we will not treat this state because we are focusing only on the origin of magnetism of 
$sp$-electrons, the planar configurations are in the ground state and can already be used to reveal 
the origin of the magnetism in nonmagnetic materials. 

We will not perform calculations for the exchange energy based on any conventional magnetic 
models. This is because the vacancy concentration in proton-irradiated graphene, in 
which ferromagnetism has been observed, corresponds to a large distance between vacancies. A 
suitable simulation for this system requires a large supercell and is computationally intensive. In 
addition, such a long-range coupling is beyond any current magnetic models. The observed 
magnetization in the proton-irradiated graphene is three or four orders of magnitude 
smaller than that of conventional magnets but with a high transition temperature (higher than room 
temperature). This is very difficult to be understood using the conventional magnetic theory. In 
other words, this type of ferromagnetism must have a distinctly different mechanism than that of 
conventional ferromagnetism. Significant conclusions could not be obtained from calculations 
based on the conventional magnetic theory; therefore, according to the nature of the nonbonding 
states, we did not perform calculations but rather postulated a possible ferromagnetic mechanism 
that is distinctly different from conventional mechanisms. 

\subsection{Graphene with a vacancy in a 2$\times$2-sized cell}
\label{sec-g-c7}

\begin{figure}[bt]
\centerline{\includegraphics[scale=0.25,angle=0]{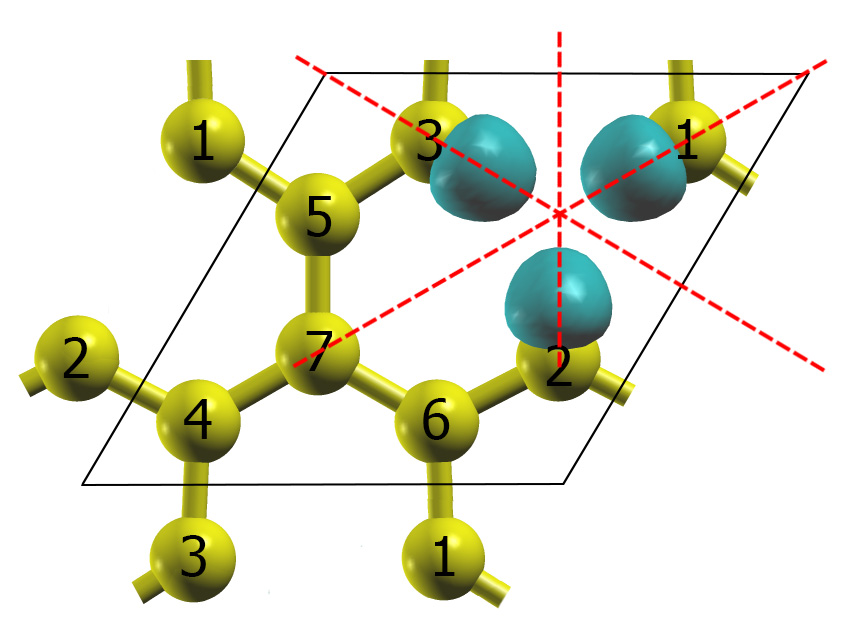}}
\caption{(Color online)
Ball and bond model for graphene with a vacancy in a 2$\times$2-sized unit cell whose 
boundary is indicated by the solid lines. The three dashed lines represent the three mirror planes 
for the unit cell without John-Teller distortion. The gray region schematically indicates charge 
distribution of the $sp^2$-type dangling bonds on atoms 1, 2 and 3.
}
\label{config}
\end{figure}

It is instructive to start the analysis of graphene in 2$\times$2-sized cell with a single-atom 
vacancy, because its magnetism is almost entirely derived from a $p_z$ state near the Fermi level, 
i.e. 0.9~$\mu_{\rm B}$ of the total 1.0~$\mu_{\rm B}$ per vacancy, which is favorable to 
revealing the origin of  the magnetism within $sp$-electron materials. The main magnetic 
contribution for the other cases (3$\times$3 $\sim$ 8$\times$8) switches from the $p_z$ states to an 
$sp^{2}$ state of the dangling bonds. However, for the 2$\times$2 case, the $sp^{2}$ state 
contributes little to the magnetism, 0.1~$\mu_{\rm B}$ of 1.0~$\mu_{\rm B}$. 

The atomic structure of the 2$\times$2-sized cell of seven carbon atoms, $g$$-$${\rm C}_7$, is 
shown in Fig.~\ref{config}. Atoms surrounding the vacancy for the other cases (3$\times$3$\sim$8$\times$8) are 
similar but with a Jahn-Teller distortion for the 4$\times$4-sized cell and larger. The atomic 
arrangement away from the vacancy can be regarded as a structural extension of perfect graphene.

\begin{figure}[bt]
\centerline{\includegraphics[scale=0.40,angle=0]{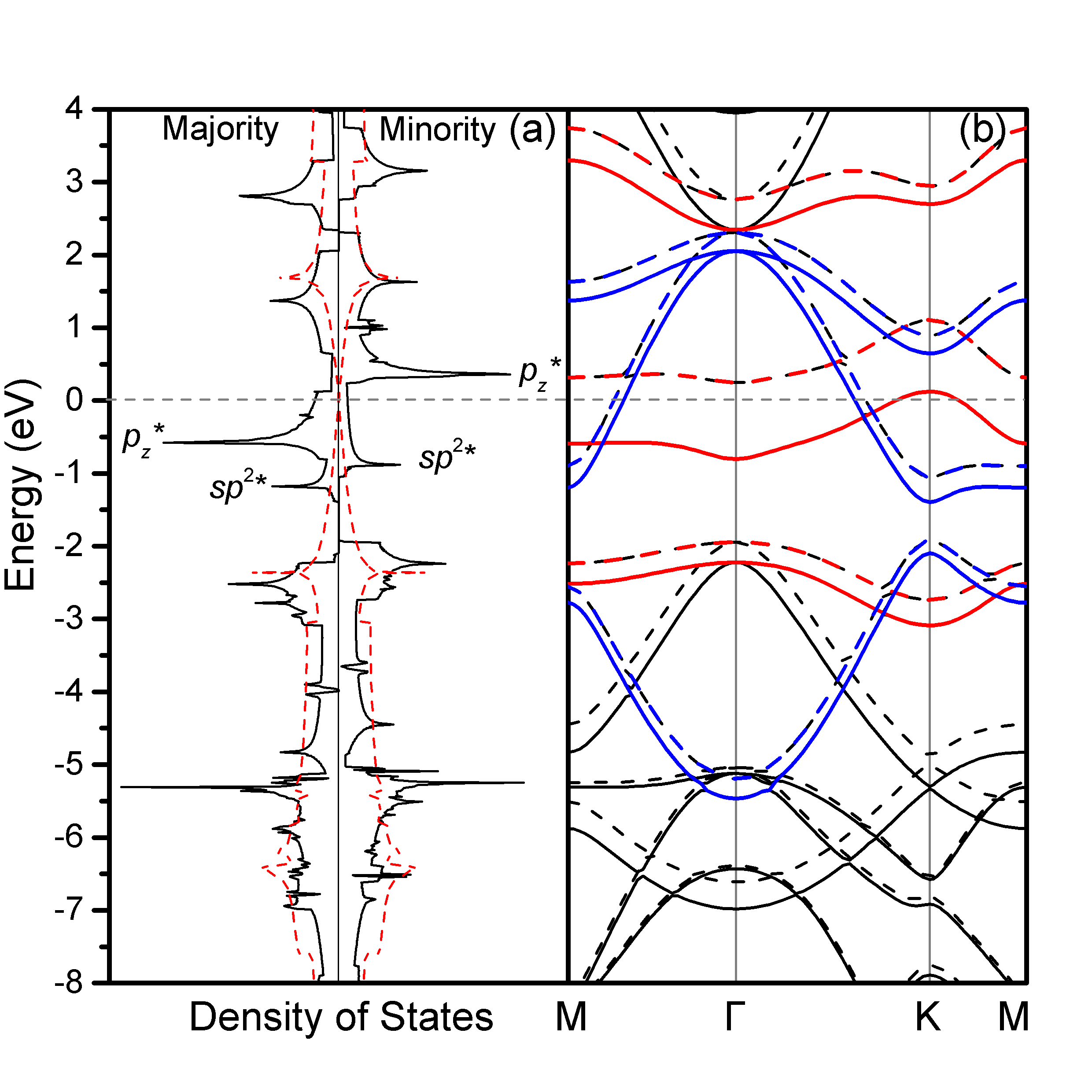}}
\caption{(Color online)
(a) DOS and (b) band structures of $g$-${\rm C}_7$. Majority and minority bands in panel (b) 
are represented by solid and dashed lines, respectively. The MM involved in the blue and red
bands in panel (b) consist of $sp^2$ and $p_z$ orbitals, respectively, labeled by $sp^{2*}$ and 
$p_z^*$ in panel (a), respectively. The Fermi level is set to zero. For comparison, the DOS for 
perfect graphene in the same sized cell are shown by dashed curves in panel (a). 
}
\label{g-c7}
\end{figure}

The DOS and band structures for $g$-${\rm C}_7$ are shown in Figs.~\ref{g-c7} (a) and ~\ref{g-c7}(b), respectively. Due 
to the planar atomic structure, the $p_z$ states and the $sp^{2}$ hybridized states are orthogonal 
without interaction so that the $p_z$ and $sp^{2}$ orbital bands can be identified in Fig.~\ref{g-c7} (b) by 
the colors red and blue, respectively; those contributing to the magnetism are labeled $p_z^*$ and 
$sp^{2*}$, respectively, in Fig.~\ref{g-c7} (a). The orbital types of these states were obtained via our 
orbital analysis based on the wavefunction projection method~\cite{vasp}, as well as on the 
maximally localized Wannier functions~\cite{Wan90}. In Fig.~\ref{g-c7} (b), the majority and minority 
bands are shown by the solid and dashed curves, respectively. For comparison, the DOS for 
graphene without a vacancy in the same sized cell as $g$-${\rm C}_7$ is given by the red dashed 
curves in Fig.~\ref{g-c7} (a). 

Comparing the two DOSs (the black solids and red dashed lines) in Fig.~\ref{g-c7} (a), the electronic 
structures of graphene near the Fermi level are shown to be largely distorted by the vacancy, 
leading to the magnetism in $g$-${\rm C}_7$. The typical features (in the energy regions $-4$~eV and 
+4~eV) of the electronic structures of graphene, which are characterized by the so-called Dirac 
point formed by the contact of the $\pi$- and $\pi^*$-bands near the Fermi level\cite{Net09}, 
disappear due to the removal of one atom (the vacancy). Instead, three red and three blue bands as 
defect states appear in this energy region. The magnetism of $g$-${\rm C}_7$ can be traced to these 
defect states near the Fermi level. The $sp^{2*}$ peaks for the majority and minority electrons 
with small spin splitting (0.3~eV) lie below the Fermi level, and only the tails of the majority and 
minority near the Fermi level show a small difference, contributing only 0.1~$\mu_{\rm B}$ of 
1.0~$\mu_{\rm B}$ per vacancy. Conversely, the spin splitting of the red state bands near the 
Fermi level is approximately 1~eV, and its majority part is nearly fully below the Fermi level, only 
a small party near the K point is above the Fermi level, while its minority part is above the Fermi 
level and unoccupied. Consequently, they contribute the majority of the MM for $g$-${\rm C}_7$, 
0.9~$\mu_{\rm B}$ of 1.0~$\mu_{\rm B}$.

\begin{figure}[bt]
\centerline{\includegraphics[scale=0.18,angle=0]{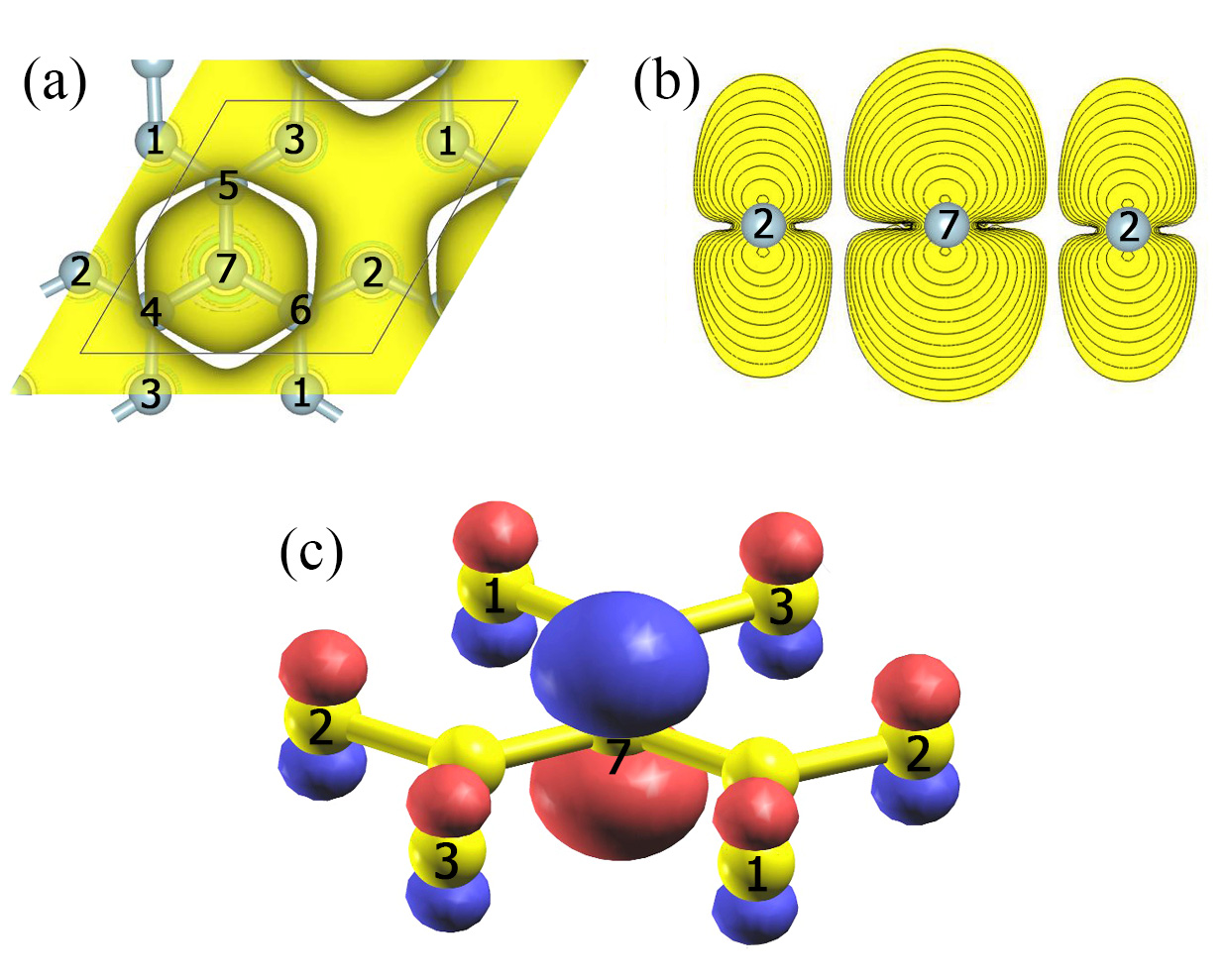}}
\caption{(Color online)
Charge distribution for the red band near the Fermi level of Fig.~\ref{g-c7} (b) at $\Gamma$. (a) an 
isosurface as a topview; (b) a contour plot of the sideview for the vertical plane cutting along the 
line linking atoms 2-7-2 in (a)£»and (c) a schematic description for the antisymmetric orbital 
combination of the state on atoms 1, 2, 3 and 7 (lobe colors up-red and down-blue, as well as 
up-blue and down-red, indicate the opposite phases of the $p_z$ orbital). Yellow balls and stick 
indicate atoms and $sp^{2}\sigma$ bonds, respectively.
}
\label{phase-g-c7}
\end{figure}

Our orbital analysis indicates that the red band near the Fermi level primarily consists of the $p_z$ 
orbitals of atoms 1, 2, 3 and 7, which belong to the same 
sublattice of graphene (referred to as sublattice A or the majority sublattice), while the removed 
atom (vacancy) belongs to the other sublattice (referred to as sublattice B or the minority 
sublattice), as do atoms 4, 5 and 6, which contribute little to the red states near the Fermi level. 
According to the symmetry of the atoms surrounding the vacancy, as shown in Fig.~\ref{config}, the atoms 
contributing to the red band can be divided into two atom groups: atom 7 (A7) and atoms 1, 2 and 
3 (A123). The orbital analysis also indicates that the red band consists of the two atom groups (A7 
and A123) in an antisymmetric manner with opposite phases as shown by the colors 
red and blue in Fig.~\ref{phase-g-c7} (c). Clearly, a node plane exists thus between the two atom groups, as shown 
by the contour plots in Figs.~\ref{phase-g-c7} (a) and \ref{phase-g-c7}(b). 
In the literature this state (the red band near the Fermi 
level) is referred to as the zero mode, implying a zero binding energy~\cite{Per06,Per08}, while 
in molecular orbital theory (MOT), it is referred to as the nonbonding state\cite{Vol11}, because 
its eigenvalue equals to the energy of the isolated $p_z$ orbital, denoted $p_z^*$ in the following. 
The $p_z^*$ state consisting of the $p_z$-orbitals of the two atom groups unexpectedly has no 
binding energy.  

Due to its importance, let us take a closer look at the unexpected nonbonding state to illustrate its 
properties. For simplicity, a tight-binding method (TB) is applied to $g$-${\rm C}_7$. In the TB model, each 
atom has only one $p_z$ orbital and only the first nearest neighbor (1NN) interaction is taken into 
account. Its Hamiltonian at $\Gamma$ is therefore
\begin{eqnarray*}
\label{eq-H7}
\left( 
\begin{array}{ccccccc}
E_0 & 0 & 0 & 0 & -t & -t & 0\\
0 & E_0 & 0 & -t & 0 & -t & 0\\
0 & 0 & E_0 & -t & -t & 0 & 0\\
0 & -t & -t & E_0 & 0 & 0 & -t\\
-t & 0 & -t & 0 & E_0 & 0 & -t\\
-t & -t & 0 & 0 & 0 & E_0 & -t\\
0 & 0 & 0 & -t & -t & -t & E_0
\end{array} 
\right).
\end{eqnarray*}
Here $E_0$ and $t$ are an on-site energy of the $p_z$ orbital and 
a hopping energy between the nearest neighbors, respectively.
The eigenvalue of the nonbonding state is $E_{\rm nonbonding}=E_0$, and the eigenfunction is 
(0.5, 0.5, 0.5, 0.0, 0.0, 0.0, -1.0) ignoring the normalization coefficients. That is, it is a state with zero 
binding energy, consisting only of the orbitals of the atom-groups A123 and A7 in an 
antisymmetric manner, without any contribution from atom-group B456, indicating that the 
nonbonding state is created by unpaired $p_z$ orbitals of atom-groups A123 and A7, not by a 
direct interaction between them because only the 1NN interaction was considered. Even though 
the state has a zero binding energy, occupying the nonbonding state still tends to stabilize the 
system as a way for electrons to synchronously appear in the two atom groups. 

Most importantly, the nonbonding state consists of orbitals from A123 and A7 in an antisymmetric 
manner. This antisymmetric combination can be understood in the following way: if the 
interaction exists only between the nearest neighbors, no direct hopping exists between A123 and 
A7. Because there is direct hopping between A123 and B456, as well as between A7 and B456, 
the interaction for the nonbonding state could be thought of a hopping bridged via B456. 
Therefore, the hopping between A123 and A7 with a zero binding energy can be realized via the 
B456 bridge. We call this bridge hopping. However, because there are no real electrons on B456 
for the nonbonding state, the wavefunction of the bridge hopping state should consist of orbitals 
from two involved atom-groups in an antisymmetric manner. This is the physics behind a state without 
any direct interaction but with an antisymmetric constraint for its wavefunction, as shown in 
Figs.~\ref{phase-g-c7} 
(b) and \ref{phase-g-c7}(c). Note that the nonbonding state here should be distinguished from a lone pair state of 
a dangling bond. A lone pair state of a dangling bond does not involve with any atoms except for 
itself. However, the nonbonding states involve at least atom-groups A123 and A7, leading therefore to an 
antisymmetric combination of the orbitals of the involved atoms.

We have seen that the MM in $g$-${\rm C}_7$ should primarily be traced to the red band (the 
$p_z^*$ nonbonding state) with a small component traced to the blue band ($sp^{2*}$) near the 
Fermi level. The $sp^{2*}$ state will be discussed in more detail in the next subsection; here we 
list its main features. The orbital analysis indicates that the $sp^{2*}$ state arises from the 
interaction between the $sp^{2}$ dangling bonds of the three atoms surrounding the vacancy. The 
three bands with the features of the $sp^{2}$ dangling bonds are identified by the color blue in 
Fig.~\ref{g-c7} (b). Only one of the three is a bonding state band, whose majority DOS peak 
lies at approximately 
$-2.8$~eV, and the other two are antibonding states whose majority DOS peaks are at approximately 
$-1.2$~eV and $+1.4$~eV. The two antibonding state bands degenerate at $\Gamma$. The three 
blue bands have dispersions of approximately 3.4~eV, 3.5~eV and 1.4~eV.
Therefore, their DOS peaks are not very sharp. 

It is well known that the nonbonding state and the antibonding state consist of antisymmetric 
orbitals. Thus, the MM of the $sp$ electrons in $g$-${\rm C}_7$ can be understood. According to 
the electron exchange antisymmetric principle of quantum mechanics, if a spatial wavefunction of a state is 
antisymmetric, its spin component should be symmetric~\cite{QM}. It has been debated for a long 
time if $sp$ electrons could lead to MM in materials. The results of $g$-${\rm C}_7$ show that the 
MM in $g$-${\rm C}_7$, created by $sp$-electrons, can be traced to fundamental principle of 
quantum mechanics. This is the physics behind the magnetism of $sp$-electrons: it is not 
$sp$-orbitals themselves but the antisymmetric behavior of their spatial wavefunctions that leads to 
electrons in spin polarization. This mechanism is different from the magnetic origin of 
$df$-electrons as proposed by Heisenberg~\cite{Hei28}, we will discuss in more detail in 
Subsection~\ref{sec-mechanism}.   

\subsection{Graphene with a vacancy in a 3$\times$3-sized cell}
\label{sec-g-c17}

\begin{figure}[bt]
\centerline{\includegraphics[scale=0.30,angle=0]{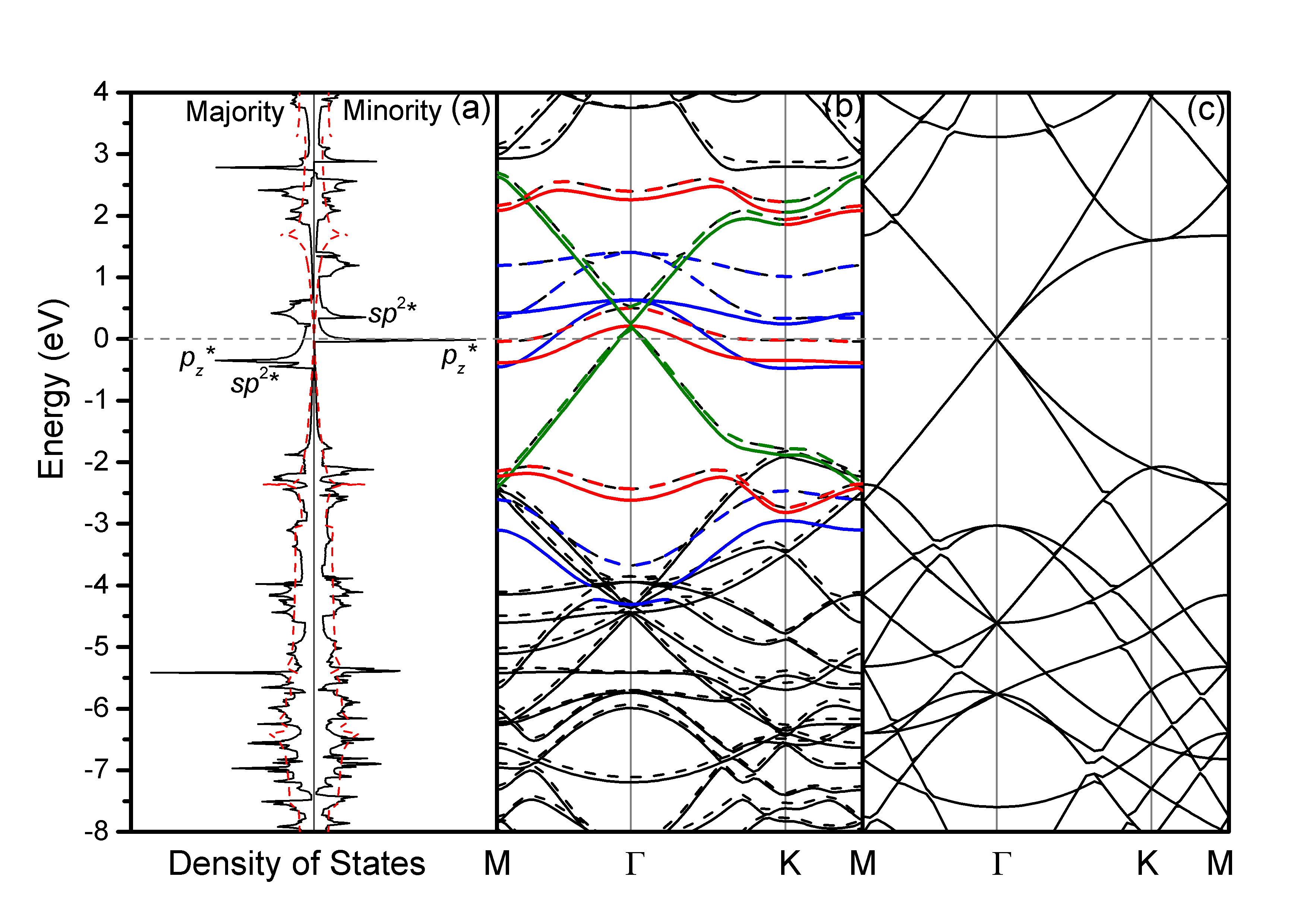}}
\caption{
(a) and (b) Same as that of Fig.~\ref{g-c7} but for $g$-${\rm C}_{17}$. (c) the band structure of the 
graphene in the same sized cell for comparison. The features with the green bands in panel (b) 
appear in all $3n$$\times$$3n$-sized cells of graphene with a vacancy, because the bands are nearly 
unaffected by the removed atom, see text for details.
}
\label{g-c17}
\end{figure}

The calculated DOS and band structures for graphene with a vacancy in a 3$\times$3-sized cell, 
$g$-${\rm C}_{17}$, are shown in Figs.~\ref{g-c17} (a) and \ref{g-c17}(b), respectively. 
Again the red and blue bands 
indicate the states composed of the $p_z$ and $sp^{2}$ orbitals, respectively. For comparison, the 
band structures of perfect graphene in the same sized cell, $g$-3$\times$3, are also shown in Fig.~\ref{g-c17} 
(c); the structures have no spin polarization as expected. 

At first glance, two green bands near the Fermi 
level in Fig.~\ref{g-c17} (b) appear to be copied from the corresponding bands of $g$-3$\times$3 in Fig.~\ref{g-c17} (c) 
and one Dirac point remains. That is, these two bands ($\pi$ and $\pi^*$) of perfect graphene are 
nearly unperturbed by the removal of an atom (the vacancy) and contact each other to form a 
Dirac point. This is entirely different from the case of $g$-${\rm C}_7$. In $g$-${\rm C}_7$, the 
vacancy removes the two Dirac points (at K and K'), leading to a formal gap near the Fermi level; 
as we discussed, all states of $g$-${\rm C}_7$ in the formal gap region can be traced to the 
vacancy-induced defect states. Clearly, the one Dirac point (formed by the two green bands) in Fig.~\ref{g-c17}
 (b) is not affected by the removal of an atom. 

We know that the Dirac points~\cite{Net09} at K and K' for $g$-1$\times$1 disappear in 
$g$-${\rm C}_7$ because the vacancy disturbs the network of the $p_z$-orbitals of graphene, therefore 
destroying the Dirac points at K and K'. It is natural to expect this feature (the disappearance of 
the Dirac points) to remain when such cells are expanded. However, it is not true for the cases of 
$3n$$\times$$3n$-sized cells (where $n$ is an integer). The natural question is, why not? 

\begin{figure}[bt]
\centerline{\includegraphics[scale=0.25,angle=0]{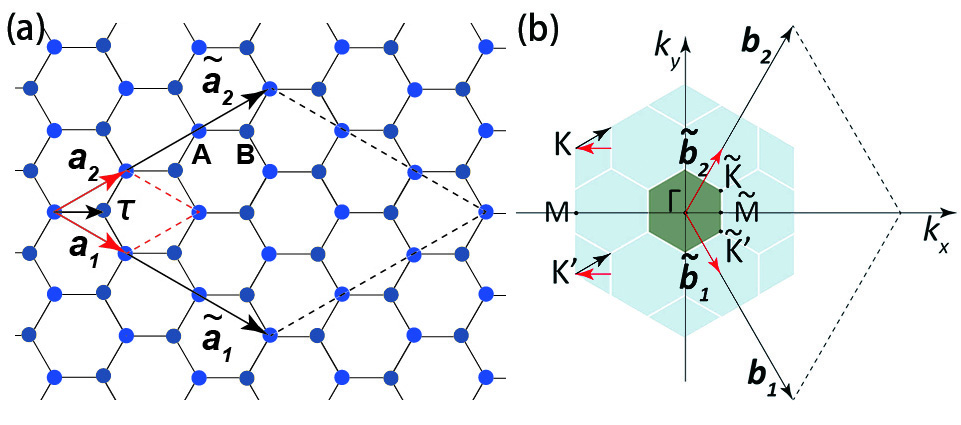}}
\caption{
The relationship of (a) the unit cells and (b) the Brillouin zones for the 1$\times$1- and 
3$\times$3-sized cells. Vectors $\mathbf{a}$ and $\mathbf{b}$ are basis vectors for the lattice and 
the reciprocal lattice, 
respectively. The tilde on the symbols indicate values for the 3$\times$3-sized cell.
}
\label{BZ}
\end{figure}

We then take $g$-${\rm C}_{17}$ as an example to answer this question. In $g$-3$\times$3, the 
Dirac point (the two bands contact each other) appears not at K and K', but degenerated at the 
$\Gamma$ point of the 3$\times$3 BZ. This is expected because of the so-called BZ folding. The 
relationship of the unit cells and BZs of graphene between the 1$\times$1- and 3$\times$3-sized 
cells is shown in Figs.~\ref{BZ} (a) and \ref{BZ}(b), respectively. Clearly, the K and K' points of the 1$\times$1 
BZ are at the $\Gamma$ point of the 3$\times$3 BZ.

The folding relationship can be also obtained by considering the relationship of the primitive 
vectors between the 3$\times$3- and 1$\times$1-sized cells. The corresponding primitive vectors 
of the 3$\times$3-sized cell,  
$\widetilde{\mathbf{a}}_1$ and $\widetilde{\mathbf{a}}_2$, 
are three times those of the 1$\times$1-sized cell,   
$\mathbf{a}_1$ and $\mathbf{a}_2$, i.e., 
$\widetilde{\mathbf{a}}_1=3\mathbf{a}_1$, 
and
$\widetilde{\mathbf{a}}_2=3\mathbf{a}_2$. 
Therefore, the relationship of the primitive vectors of the 
reciprocal lattice between the two lattices should be
$\widetilde{\mathbf{b}}_1=1/3\mathbf{b}_1$ and 
$\widetilde{\mathbf{b}}_2=1/3\mathbf{b}_2$. 
The K and  K' points for the 1$\times$1-sized cell can be represented by 
$\mathbf{b}_1$  and $\mathbf{b}_2$  as 
$\mathbf{K}=2/3\mathbf{b}_1+1/3\mathbf{b}_2$  and
$\mathbf{K}'=1/3\mathbf{b}_1+2/3\mathbf{b}_2$. 
Therefore, by means of the primitive vectors of 
the reciprocal lattice for the 3$\times$3-sized cell, 
$\widetilde{\mathbf{b}}_1$  and  $\widetilde{\mathbf{b}}_2$, 
they should be written as 
$\mathbf{K}=2/3\mathbf{b}_1+1/3\mathbf{b}_2=2\widetilde{\mathbf{b}}_1+\widetilde{\mathbf{b}}_2$ and  
$\mathbf{K}'=1/3\mathbf{b}_1+2/3\mathbf{b}_2=\widetilde{\mathbf{b}}_1+2\widetilde{\mathbf{b}}_2$. 
$\mathbf{K}$ and  $\mathbf{K}'$ are the two 
reciprocal lattice vectors for the 3$\times$3-size cell. One end-point of a reciprocal lattice vector 
is the center of its Brillouin zone, the $\Gamma$ point. That is, both K and K' of the $1\times 
1$-sized cell are folded to and degenerated at one point, the $\Gamma$ point of the $3\times 
3$-sized cell. This is also valid for all cells with sizes of $3n$$\times$$3n$. 

The Dirac point for $g$-${\rm C}_{17}$ is fourfold degenerated because both the K and K' point of 
the 1$\times$1 BZ are folded to the $\Gamma$ point of the 3$\times$3 BZ. Furthermore, it can 
be seen from Fig.~\ref{g-c17} (c) that, along the M-$\Gamma$ axis, the bands of $g$-3$\times$3 just below 
and above the Fermi level are degenerated while along the $\Gamma$-K axis, the degeneration of 
the bands is lifted because the two axes pointed by two red arrows 
(from $\widetilde{{\rm M}}$ to $\widetilde{\Gamma}$ of the 
3$\times$3 BZ) are equivalent in the 1$\times$1 BZ, while that of the two black arrows (from 
$\widetilde{\Gamma}$ to $\widetilde{{\rm K}}$ of the 3$\times$3 BZ) are not, as shown in Fig.~\ref{BZ} (b). 

Keeping this relationship in mind, we can now understand why the one Dirac point does not 
disappear even if one atom is removed from $g$-3$\times$3 and two green bands in Fig.~\ref{BZ} (b) 
resemble the corresponding bands in $g$-3$\times$3. For this discussion, we again use the TB 
model. In the 1$\times$1-sized cell of graphene, the Bloch sums at sites A and B (labeled in Fig.~\ref{BZ}  
(a)) are
\begin{eqnarray*}
\psi^{A}_{\bf{k}}=\sum_{\bf{R}}\phi^A\left(\bf{r}-\bf{R}\right)e^{i\bf{k}\cdot \bf{R}}
\end{eqnarray*}
and
\begin{eqnarray*}
\psi^{B}_{\bf{k}}=\sum_{\bf{R}}\phi^B\left(\bf{r}-\bf{R}-\bm{\tau}\right)
  e^{i\bf{k}\cdot\left(\bf{R}+\bm{\tau}\right)},
\end{eqnarray*}
respectively. They form the so-called $\pi$ and $\pi^*$ states in the symmetric and antisymmetric 
types, respectively, (ignoring the normalization coefficients) as
\begin{widetext}
\begin{eqnarray*}
\psi^{\pi}_{\bf{k}}=\sum_{\bf{R}}\phi^A\left(\bf{r}-\bf{R}\right)
  e^{i\bf{k}\cdot \bf{R}}+C_{\bf{k}}\sum_{\bf{R}}
  \phi^B\left(\bf{r}-\bf{R}-\bm{\tau}\right)
  e^{i\bf{k}\cdot\left(\bf{R}+\bm{\tau}\right)}
\end{eqnarray*}
\end{widetext}
and
\begin{widetext}
\begin{eqnarray*}
\psi^{\pi^\ast}_{\bf{k}}=\sum_{\bf{R}}\phi^A\left(\bf{r}-\bf{R}\right)
  e^{i\bf{k}\cdot \bf{R}}-C_{\bf{k}}\sum_{\bf{R}}
  \phi^B\left(\bf{r}-\bf{R}-\bm{\tau}\right)e^{i\bf{k}\cdot\left(\bf{R}+\bm{\tau}\right)}.
\end{eqnarray*}
\end{widetext}
Because the 3$\times$3-sized cell is only artificially extended from 1$\times$1-sized cell, the 
bands' properties (bonding and antibonding) should be maintained, the $\pi$ and $\pi^*$ bands 
along the M-$\Gamma$ axis for $g$-3$\times$3 should still be composed of the $\pi$ and $\pi^*$ 
in $g$-1$\times$1. For the BZ relationship between $g$-1$\times$1 and $g$-3$\times$3, please refer to 
Fig.~\ref{BZ}.  $\bm{\tau}$ has no $y$-component, and   because $\bf{k}$ and $\bf{k}$' in $g$-1$\times$1 are 
symmetric in the $x$-direction, as shown in Fig.~\ref{BZ} (b); therefore, along the M-$\Gamma$ of 
$g$-3$\times$3, we obtain
\begin{widetext}
\begin{eqnarray*}
\begin{array}{ll}
\Psi^{\pi}_{A} 
  &=\psi^{\pi}_{\bf{K}+\Delta\bf{k}}-\psi^{\pi}_{\bf{K'}+\Delta\bf{k}}\\
  &=\sum\limits_{\bf{R}}\phi^A\left(\bf{r}-\bf{R}\right)
    e^{i\Delta\bf{k}\cdot\bf{R}}\left(e^{i\bf{K}\cdot\bf{R}}
    -e^{i\bf{K}'\cdot\bf{R}}\right)\\
  & +\sum\limits_{\bf{R}}e^{i\left(\frac{4\pi}{3}+\Delta k_x a\right)}
    \phi^B\left(\bf{r}-\bf{R}-\bm{\tau}\right)
    e^{i\Delta\bf{k}\cdot\bf{R}} \left[C_{\bf{K}+\Delta\bf{k}}
    e^{i\bf{K}\cdot\bf{R}}-C_{\bf{K}'+\Delta\bf{k}}
    e^{i\bf{K}'\cdot\bf{R}}\right]\\
  & =\sum\limits_{\bf{R}\left(\ne \bf{R}'\right)}\left[\phi^A\left(\bf{r}-\bf{R}\right)
    +C_{\bf{K}+\Delta\bf{k}}e^{i\left(\frac{4\pi}{3}+\Delta k_x a\right)}
    \phi^B\left(\bf{r}-\bf{R}-\bm{\tau}\right)\right] 
    e^{i\Delta\bf{k}\cdot\bf{R}}\left(e^{i\bf{K}\cdot\bf{R}}
   -e^{i\bf{K}'\cdot\bf{R}}\right)
\end{array}
\end{eqnarray*}
\end{widetext}
and
\begin{widetext}
\begin{eqnarray*}
\begin{array}{ll}
\Psi^{\pi^\ast}_{A}
  &=\psi^{\pi^\ast}_{\bf{K}+\Delta\bf{k}}-\psi^{\pi^\ast}_{\bf{K}'+\Delta\bf{k}}\\
  &=\sum\limits_{\bf{R}\left(\ne \bf{R}'\right)}\left[\phi^A\left(\bf{r}-\bf{R}\right)
    -C_{\bf{K}+\Delta\bf{k}}e^{i\left(\frac{4\pi}{3}+\Delta k_x a\right)}
    \phi^B\left(\bf{r}-\bf{R}-\bm{\tau}\right)\right] 
    e^{i\Delta\bf{k}\cdot\bf{R}}\left(e^{i\bf{K}\cdot\bf{R}}-
    e^{i\bf{K}'\cdot\bf{R}}\right)
\end{array}
\end{eqnarray*}
\end{widetext}
with  $C_{\mathbf{K}+\Delta\mathbf{k}}=C_{\mathbf{K}'+\Delta\mathbf{k}}$. 
The above sum of $\mathbf{R}$ excludes $\mathbf{R}'$ because 
$\mathbf{K}\cdot \mathbf{R}=\mathbf{K}'\cdot \mathbf{R}+2\pi\left(n_1-n_2\right)/3$
for all $\mathbf{R}'$  lattice vectors satisfying $n_1-n_2=3n$  with $n$ as 
an integer and  
$e^{i\mathbf{K}\cdot\mathbf{R}}-e^{i\mathbf{K}'\cdot\mathbf{R}}=0$. 
Therefore, the two-fold degenerate bands along the 
M-$\Gamma$ axis of $g$-3$\times$3 are independent of all lattice vectors $\mathbf{R}'$. 
Therefore, removing 
one atom from graphene (creating a vacancy) in a $3n$$\times$$3n$-sized cell will not significantly affect 
the one 
Dirac point and the involved bands (green) if it is modeled using the TB model. 

Note that the green bands have nearly no spin polarization, as shown in Fig.~\ref{g-c17} (b). Like the red 
bands near the Fermi level, the green bands also consist of the $p_z$ orbitals and also appear in 
the same energy region. Therefore, the reason that the green bands do not contribute to magnetism 
but the red bands do is not related to the intrinsic nature of the $p_z$ orbitals but to the 
antisymmetric type of the wavefunctions, indicating that, unlike $df$ electrons, $sp$ electrons 
have only small amounts of exchange energy. No defect states (antibonding states or nonbonding 
states here) mean no magnetism for $sp$ electrons in an $sp$-bonded crystal. This may be why 
$sp$-electron magnetism has been in doubt for such a long 
time~\cite{Kat12,Kuz13,Han14,Fis15}. 

The MM of $g$-${\rm C}_7$ are primarily derived from the $p_z^*$ nonbonding states, while the 
main magnetic contributor in $g$-${\rm C}_{17}$ is the $sp^{2*}$ blue bands near the Fermi level. 
Now, we focus on the blue bands in $g$-${\rm C}_{17}$. The orbital analysis indicates that the blue 
bands consist of $sp^{2}$-type orbitals of the three C atoms surrounding the vacancy, atoms 1, 2, 
and 3. From our previous investigations of $g$-${\rm C}_{14}{\rm N}_3$~\cite{Yu161,Yu162} and 
$g$-${\rm C}_4{\rm N}_3$~\cite{Xu17}, it is known that the dangling bonds of the three N atoms interact each other 
to form one bonding state and two antibonding states. The difference between $g$-${\rm C}_{14}{\rm N}_3$ ($g$-${\rm C}_4{\rm N}_3$) 
and $g$-${\rm C}_{17}$ ($g$-${\rm C}_{7}$) is that the three N atoms surrounding the vacancy in $g$-${\rm C}_{14}{\rm N}_3$
($g$-${\rm C}_4{\rm N}_3$) are replaced by three C atoms. The two antibonding states, degenerated at the 
$\Gamma$-point, have a mirror-symmetric combination (MSC) and a mirror-antisymmetric 
combination (MAC) of the $sp^{2}$ orbitals with $C_{\rm 3V}$ symmetry. Examining $g$-${\rm C}_{17}$ 
($g$-${\rm C}_{7}$), the situation is similar to $g$-${\rm C}_{14}{\rm N}_3$ ($g$-${\rm C}_4{\rm N}_3$)~\cite{Yu161,Yu162}: 
the three blue bands 
are also induced by the interaction between the three dangling bonds on the three atoms 
surrounding the vacancy. When one C atom is removed from graphene, the three $sp^{2}$ 
dangling bonds on the three atoms surrounding the vacancy are left. The dangling bonds, which 
were originally connected to the removed atom, are not fully filled. Therefore, they interact with 
each other and form one bonding state and two antibonding states. 

To illustrate the resulting three (one bonding and two antibonding) states, a TB model is 
again taken into account. For three atoms each with only a dangling bond orbital, the Hamiltonian 
is 
\begin{eqnarray*}
\label{eq-H3}
\left( 
\begin{array}{ccc}
E_0 & -t  & -t \\
-t  & E_0 & -t \\
-t  & -t  & E_0 \\
\end{array} 
\right).
\end{eqnarray*}
Here $E_0$ and $t$ are the energy of the dangling bond and the interaction between the dangling bonds, respectively. 
We can obtain the energy of the bonding 
state as $E_1=E_0-2t$ and those of the doublet degenerated antibonding states as $E_2=E_3=E_0+t$. The 
difference between $g$-${\rm C}_{17}$ ($g$-${\rm C}_{7}$) and $g$-${\rm C}_{14}{\rm N}_3$ ($g$-${\rm C}_4{\rm N}_3$) 
is that the MSC and MAC 
bands are partially occupied in the cases of $g$-${\rm C}_{17}$ and $g$-${\rm C}_7$, while in the 
cases of $g$-${\rm C}_4{\rm N}_3$ and $g$-${\rm C}_{14}{\rm N}_3$, both bands are nearly fully occupied. 
This is because $g$-${\rm C}_{14}{\rm N}_3$
($g$-${\rm C}_4{\rm N}_3$) has more three electrons than $g$-${\rm C}_{17}$ ($g$-${\rm C}_{7}$) and is a +1$e$ hole system, while 
$g$-${\rm C}_{17}$ ($g$-${\rm C}_{7}$) is a +4$e$ hole system. Therefore, the Fermi level is lowered relative to 
the top of the MSC and MAC bands in $g$-${\rm C}_{17}$ ($g$-${\rm C}_{7}$).

\begin{figure}[bt]
\centerline{\includegraphics[scale=0.18,angle=0]{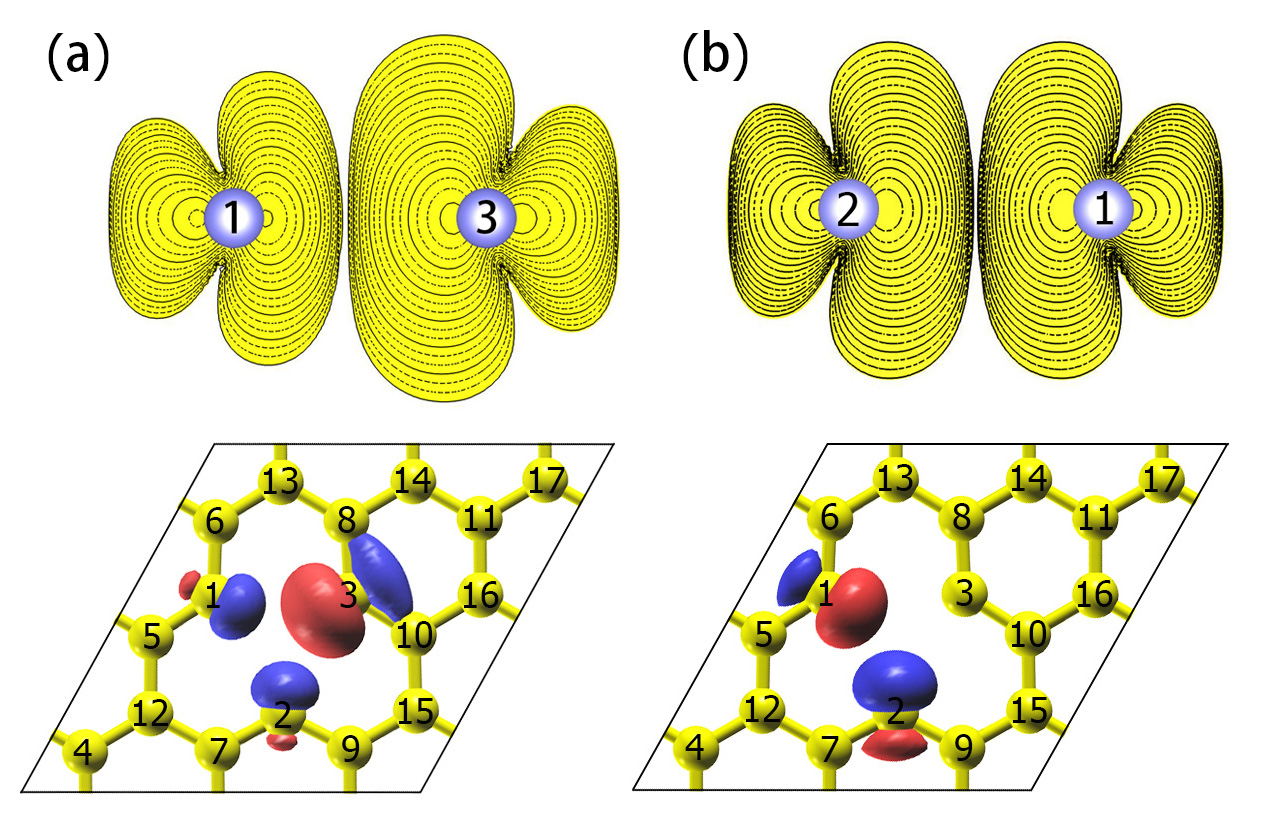}}
\caption{
Charge distribution for the red band in Fig.~\ref{g-c17} (a) at $\Gamma$ for the (a) MSC and (b) for 
MAC states. Panels (a) and (b) are the up and down panels for the side and top views, respectively. 
In the side view, the vertical plane cuts along the line linking atoms 1-3 and atoms 1-2 for 
panels (a) and (b), respectively.
}
\label{phase-g-c17}
\end{figure}

The features of the MSC and MAC antibonding states in the charge distribution at $\Gamma$ are 
shown in Figs.~\ref{phase-g-c17} (a) and \ref{phase-g-c17}(b), respectively. 
The blue and red colors in Fig.~\ref{phase-g-c17} (b), 
as that in Fig.~\ref{phase-g-c7}  
(c), indicate the different phases, and nodes exists between atoms 1(2) and 3 for the MSC state, as 
well as between atoms 1 and 2 for the MAC state, as shown by the counter plots in Figs.~\ref{phase-g-c17} (a) and 
\ref{phase-g-c17}(b), indicating an antisymmetric combination of the involved orbitals on the three atoms.  

Note that previous calculations attributed the magnetic contribution of the $sp^{2}$-type states in 
graphene with vacancies to JTD~\cite{Ma04,Yaz07,Nan12,Sin13}. However, even without 
JTD in $g$-${\rm C}_{17}$ the $sp^{2*}$ states contribute to the magnetism, indicating that the 
magnetism in graphene with vacancies is independent of JTD. The origin of the magnetism arising 
from the states can be also traced to the antisymmetric mode of their wavefunctions.  

\subsection{Graphene with a vacancy in a 4$\times$4-sized cell}
\label{sec-g-c31}
For cell sizes larger than 4$\times$4, the optimized atomic structure for graphene with a vacancy 
has a JTD. That is, two of the three atoms surrounding the vacancy move closer to each other, 
while the third one moves away, forming thus an isosceles triangle with one short and two long 
sides. The rest of the atoms in the unit cell experience only minor change due to JTD, whereas the 
relevant electronic structures are not significantly affected by the minor change.

\begin{figure}[bt]
\centerline{\includegraphics[scale=0.37,angle=0]{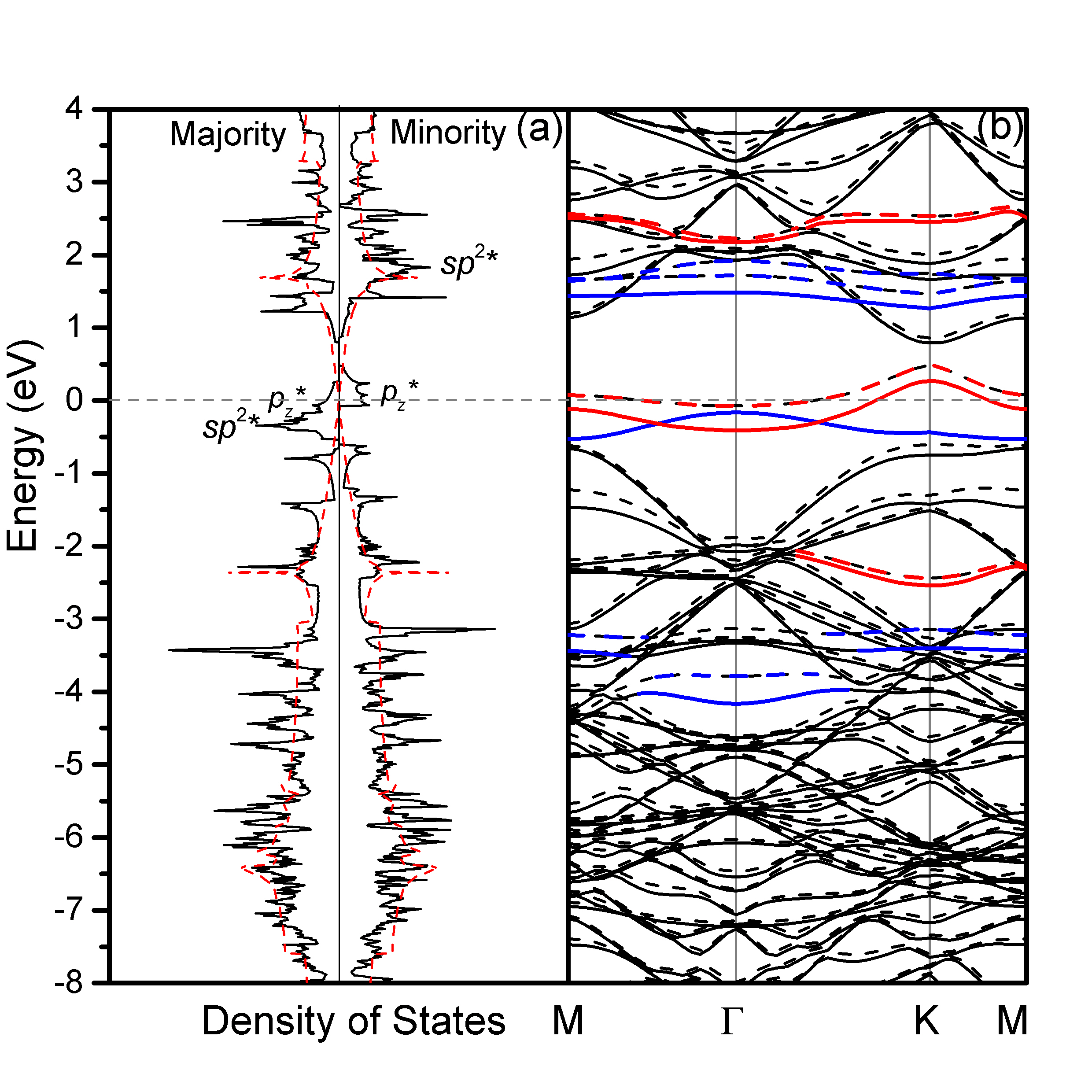}}
\caption{
The same as Fig.~\ref{g-c7} and Fig.~\ref{g-c17} but for $g$-${\rm C}_{31}$.
}
\label{g-c31}
\end{figure}

As in $g$-${\rm C}_7$ and $g$-${\rm C}_{17}$, the vacancy-induced $sp^{2}$ and $p_z$ bands 
are again shown in blue and red, 
respectively, in the band structure in Fig.~\ref{g-c31} (b). As shown in Fig.~\ref{g-c31}, the $p_z^*$-nonbonding state 
(red) bands, due to the unpaired $p_z$ orbitals of the atoms in the same sublattice, still appear 
near the Fermi level as expected. Because the $p_z$ and $sp^{2}$ orbitals are orthogonal, the 
nonbonding state bands are not significantly affected by JTD, contributing approximately 
0.6~$\mu_{\rm B}$, consistent with previous calculations (referred to as the zero mode 
states)~\cite{San10,Sun17}. As discussed in $g$-${\rm C}_7$ and $g$-${\rm C}_{17}$, note that the 
important factor for the nonbonding states is not whether the binding energy is zero but its 
antisymmetric orbital combination, to which the magnetic contribution can be traced. The 
occupation of some minority nonbonding states can be traced to the competition between the 
$\bf{k}$-dependent kinetic energy and the exchange energy. The other two $p_z$ orbital bands, namely 
the $\pi$- and $\pi^*$-bands (red), lie at approximately $-2.5$~eV and 2.5~eV respectively, and do 
not contribute to the MM because their majority and minority bands are either fully occupied or 
completely empty. 

The main contributor to the magnetism in $g$-${\rm C}_{31}$ is the blue bands near the Fermi 
level, as shown in Figs.~\ref{g-c31} (a) and \ref{g-c31}(b); 
the majority blue band near the Fermi level is fully occupied, 
while the corresponding minority one ($\sim$ 1.8~eV) is completely empty, therefore contributing 
1.0~$\mu_{\rm B}$ to the magnetic moment of $g$-${\rm C}_{31}$. The JTD parameters (long side 
length 2.58~\AA~ and short side 2.18~\AA), the band dispersion and the contributing magnitude to the 
magnetic moment of the blue bands obtained by our calculations are in agreement with the 
corresponding calculations~\cite{Yaz07,San10,Nan12,Sun17}. However, the explanation is 
different. It was suggested~\cite{Yaz07,Nan12,Pad16} that, due to JTD, two of the three atoms 
around the vacancy form a $sp^{2}\sigma$ bond, leaving the apical atom with a dangling bond 
contributing a magnetic moment of approximately 1.0~$\mu_{\rm B}$ according to Hund's rule. 
However, because the short side is much larger than an $sp^{2}\sigma$ bond length in perfect 
graphene, 1.42~\AA~\cite{Net09}, it is impossible for the two atoms to form an $sp^{2}\sigma$ bond, 
leaving the apical atom alone to form a solitary dangling bond and to be filled by one electron 
with spin polarization. Even if a solitary dangling bond exists on one atom, 
the magnetism obtained by calculations cannot 
be traced to it because in the calculations based on the SEA, one electron can half fill on one spin 
channel and half on the other spin channel. 
A famous example contrary to the previous conclusion 
that the magnetism in graphene with vacancies is contributed to the dangling bond is the dangling 
bonds on the Si(111) surface. Dangling bond states exist on the Si(111) surface and their electron 
states satisfy the local condition. However, the Si(111) surface does not show any spin 
polarization~\cite{Si111} because a half electron occupies the spin-up channel and the other half 
one can occupy the spin-down channel of the dangling bond state. 
Therefore, Hund's rule does not hold when interpreting results calculated based on SEA.  

The vacancy-induced $sp^{2*}$ states are formed by the interaction between all three atoms 
surrounding the vacancy, forming one bonding and two antibonding states and identified by the 
color blue in Fig.~\ref{g-c31} (b). Even though the $C_{\rm 3V}$ symmetry in $g$-${\rm C}_{31}$ is broken due to 
JTD, two of the three blue bands, which consist of the $sp^{2}$ orbitals of the three atoms around 
the vacancy and lie near the Fermi level and $\sim$$+1.5$~eV, respectively, can still be characterized by 
MSC and MAC. This is a key to understanding the magnetism in 
graphene with vacancies. That is, all states that contribute to the magnetism consist of 
antisymmetric spatial wavefunctions. When the states appear near the Fermi level, they are 
partially occupied in spin polarization according to the electron exchange antisymmetric principle of 
quantum mechanics. This is a different mechanism than that suggested by Heisenberg for 
$df$-electrons~\cite{Hei28}, which we will discuss later. 

To further illustrate the antisymmetric nature of MSC, we built a TB model of three $sp^{2}$ 
dangling bonds on three atoms with a JTD; only one $sp^{2}$ dangling bond for each atom of the 
three atoms was taken into account. The Hamiltonian of the TB model can be easily written as
\begin{eqnarray*}
\label{eq-2NN}
\left(
\begin{array}{ccc}
E_0 & -t_1 & -t_1 \\
-t_1 & E_0 & -t_2 \\
-t_1 & -t_2 & E_0 \\
\end{array}
\right).
\end{eqnarray*}
Here $t_1$ and $t_2$ are the interactions between two atoms on the long and short sides of an 
isosceles triangle, respectively, with $t_2>t_1$. 
One can obtain  
$E_1=E_0-\frac{1}{2}\left(t_2+\sqrt{t^2_2+8t^2_1}\right)$, 
$E_2=E_0+\frac{1}{2}\left(\sqrt{t^2_2+8t^2_1}-t_2\right)$
and  
$E_3=E_0+t_2$. 
The corresponding eigenfunctions for $E_1$, $E_2$ and $E_3$ are 
$\left(\frac{-t_2+\sqrt{8t^2_1+t^2_2}}{2t_1},1,1\right)$, 
$\left(\frac{-t_2-\sqrt{8t^2_1+t^2_2}}{2t_1},1,1\right)$ and 
$\left(0,-1,1\right)$, respectively, clearly showing 
the antisymmetric combination for $E_2$ and $E_3$. This means that, if the three unsaturated dangling 
bonds interact with each other, they would form one bonding state with a symmetric orbital 
combination and two antibonding states each with an antisymmetric orbital combination. Clearly, 
the two antibonding states, $E_2$ and $E_3$, correspond to MSC and MAC respectively, while $E_1$ 
corresponds to the $sp^{2}$ bonding state. 

Therefore, we can understand the magnetism contributed by the MSC state: JTD breaks the 
constrain of the $C_{\rm 3V}$ symmetry, therefore lifting the degeneration of MSC and MAC at $\Gamma$. 
The MSC $sp^{2*}$ band (blue) is therefore lower in energy than the $p_z^*$-nonbonding state 
due to $s$-orbital components in the $sp^{2*}$ states; the majority of the MSC band is 
fully filled by one electron, leaving its minority band completely empty and contributing 
1.0~$\mu_{\rm B}$ according to the electron exchange antisymmetric principle of the quantum 
mechanics. 

\subsection{Graphene with a vacancy in 5$\times$5 $\sim$ 8$\times$8-sized cells}
\label{sec-5-8}

\begin{figure}[bt]
\centerline{\includegraphics[scale=0.30,angle=0]{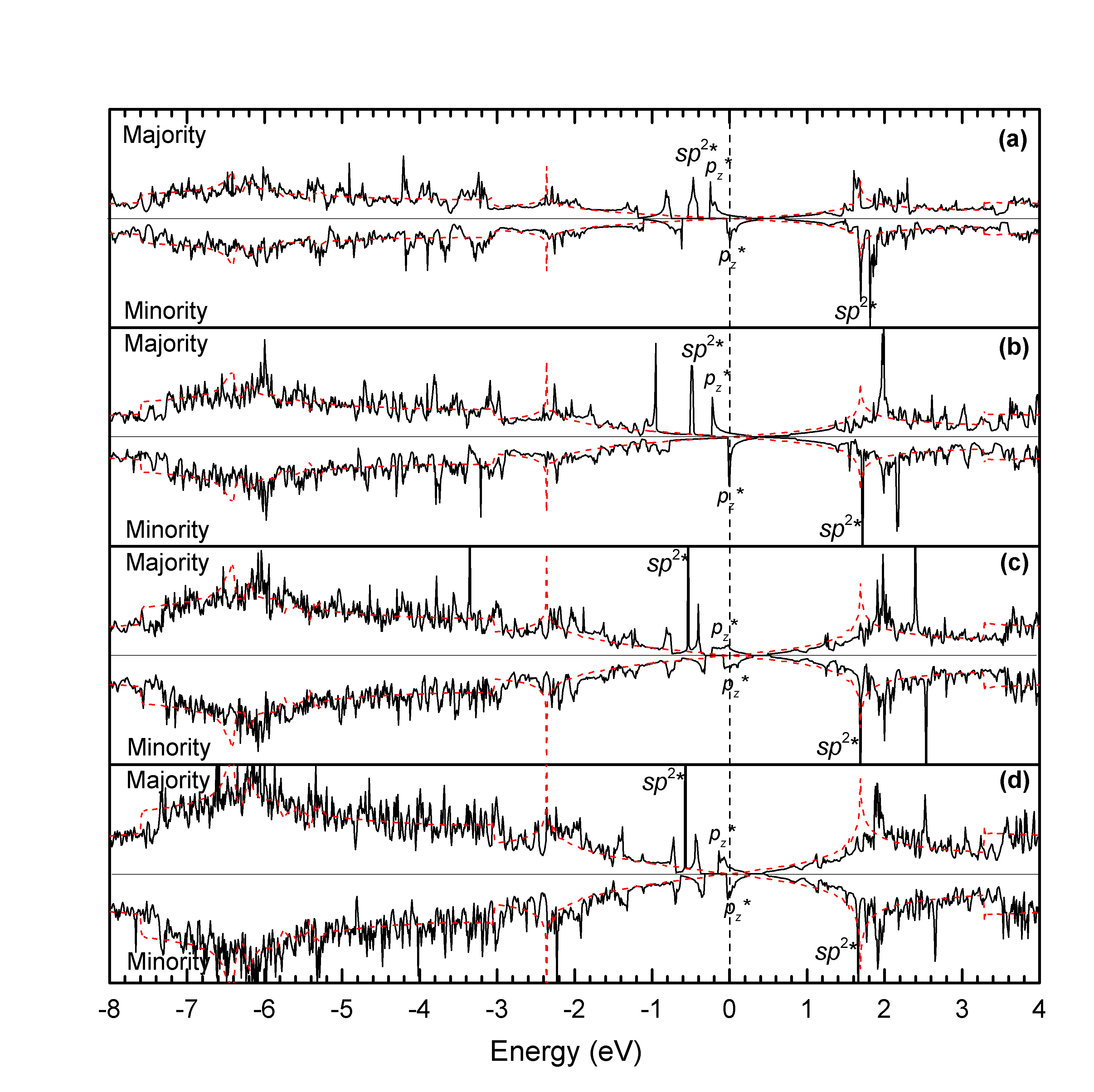}}
\caption{
Density of states for graphene with a vacancy in 5$\times$5 (a), 6$\times$6 (b), $7\times 
7$ (c) and 8$\times$8 (d) size. The symbols of $sp^{2*}$ and $p_z^*$ have the same orbital type 
as in Figs.~\ref{g-c7}, \ref{g-c17} and~\ref{g-c31}.
}
\label{dos-5-8}
\end{figure}

Electronic structures for 5$\times$5 $\sim$ 8$\times$8-sized cells of graphene with a vacancy are 
similar to the corresponding structures for the 2$\times$2 $\sim$ 4$\times$4-sized cells. Importantly, 
the MM in these cases are derived from the antibonding and nonbonding states. Here we briefly 
discuss the magnetism-concerned electronic structures in the 5$\times$5 $\sim$ 8$\times$8-sized cells 
by means of DOS, as shown in Figs.~\ref{dos-5-8} (a)$\sim$\ref{dos-5-8}(d). 

Similar to the 4$\times$4 case, the peaks of the majority antibonding states (labeled by 
$sp^{2*}$ as shown in Fig.~\ref{dos-5-8}) lie at approximately -0.6~eV for the involved cases 
(5$\times$5 $\sim$ 8$\times$8) and are fully occupied, while their minority peaks lie at approximately +1.6~eV 
and are completely empty. Therefore, the $sp^{2*}$-antibonding states contribute a magnetic 
moment of 1.0~$\mu_{\rm B}$. These peaks become increasingly sharper with increasing cell sizes, 
implying that the dispersion of the corresponding bands decrease. It is expected that the 
dispersions of the bands depend on the interaction of the $sp^{2*}$-antibonding states over the 
unit cell; this interaction is therefore increasingly weaker as the cell sizes increase. 

Conversely, the peak width of the $p_z^*$-nonbonding states, which lie around $-0.2$ and $\sim$ 0.0~eV for the 
majority and minority, respectively, and contribute to MM of 0.6$\sim$0.3~$\mu_{\rm B}$, does not 
depend as much on the cell sizes because their wavefunctions are widely extended, decaying with 
the distance $r$ to the vacancy as $\sim$ $1/r$~\cite{Nan12}. This feature can be understood because these 
states are induced by the unpaired $p_z$ orbitals, involving not only two atoms but also 
additional atoms in the majority sublattice of the unit cell. This is a very important feature that 
determines the ferromagnetism in graphene with vacancies via a never before reported mechanism, 
see the next subsection for details. From Table I, the MM for cases 5$\times$5 $\sim$ 8$\times$8 are 
1.6~$\mu_{\rm B}$, 1.6~$\mu_{\rm B}$, 1.3~$\mu_{\rm B}$ and 1.3~$\mu_{\rm B}$, respectively. 
Except for the contribution of the $sp^{2*}$-antibonding states, 1.0~$\mu_{\rm B}$, the magnetic 
moment of 0.3 $\sim$ 0.6~$\mu_{\rm B}$ can be traced to different electron occupations in the majority 
and minority of the $p_z^*$-nonbonding states. As discussed in the cases of 2$\times$2 $\sim$ $4\times 
4$, this is a result of the competition between the kinetic energy and the exchange energy. For 
cases 7$\times$7 and 8$\times$8, more minority nonbonding states are occupied, reducing the 
MM. 

\subsection{Unusual ferromagnetic mechanism distinctly different from the conventional 
mechanism} 
\label{sec-mechanism}
As mentioned above, when the origin of the magnetism in proton-irradiated 
graphene was explained by the conventional magnetic theory, it was debated or resulted in 
conflicts. The current magnetic models, such as the indirect exchange model, the superexchange 
model, the double exchange model and the RKKY model, 
have the same scalar product in the exchange energy, $\mathbf{S}_i\cdot\mathbf{S}_j$; the difference lies 
only in the manner of coupling. In this aspect, these models are all based on Heisenberg's theory 
originated from the exchange integrals for an H$_2$ molecule\cite{Hei28}. 
The magnetic mechanism for magnetic materials such as 
iron and nickel metals is governed by the conventional magnetic theory. 

If the conventional magnetic theory is used to explain the magnetism in graphene with vacancies, 
contradictions and difficulties lie at least in two aspects. 1) How can $sp$ electrons (principle 
quantum number $n<3$) be spin-polarized? 2) How can such a weak magnetization 
($10^{-3}$$\sim$ $10^{-4}$ orders of magnitude smaller than a conventional magnet) have such a high 
critical temperature? In other words, how is such a long-range (with a distance between vacancies 
of up to 20~\AA) coupling strong enough to trigger ferromagnetic ordering above room 
temperature\cite{Elf02,Uge10,Esq13}? There must be an unrecognized magnetic mechanism at 
work. 

Concerning the difficulty in explaining $sp$ electron spin-polarization in proton-irradiated 
graphene, we have shown that the vacancy-induced states such as antibonding and 
nonbonding states can appear near the Fermi level. If these states are partially filled, according to 
the antisymmetric principle of electron exchange the $sp$ electrons could be spin-polarized. 
Obviously, unlike an isolated atom governed by Hund's rule, the $sp$-electron spin polarization in 
proton-irradiated graphene originates from the antisymmetric spatial 
wavefunctions involving three atoms surrounding the vacancy for $sp^{2*}$
and all the atoms of the majority 
sublattice for $p_z^*$. The induced MM of the atoms 
are not individually localized but are distributing as a whole moment on all the involved atoms. If 
the entire magnetic moment is projected on the atoms, the projected moments appear to be 
distributed on the involved atoms. To distinguish these from localized MM,
we refer to these moments as fractional magnetic moments. The fractional MM 
on the majority sublattice atoms, whose $p_z$ orbitals are composed in an antisymmetric manner, 
are inseparably combined into the entire magnetic moment,
because they belong to one electronic state, the $p_z^*$ state.  
This is a distinctly different mechanism from any conventional mechanism. 
We used the word combine in the last sentence to emphasize that there is no 
coupling interaction between the fractional magnetic moments, 
rather the intrinsic parts belong to the entire 
magnetic moment. These fractional MM, as a whole moment induced by 
partially filled antibonding and nonbonding states, should be parallel according to the exchange 
antisymmetric principle for electrons.

Concerning the difficulty in explaining the ferromagnetic ordering with such a high critical 
temperature in proton-irradiated graphene, we have stressed in 
Subsection~\ref{summary} that we did not perform calculations for the exchange energy 
based on the conventional magnetic theory for two reasons. 1) Because the vacancy concentration 
in proton-irradiated graphene, for which ferromagnetism was observed, corresponds 
to a distance between vacancies as large as 15$\sim$20~\AA; a valid simulation would be computationally 
demanding. 2) Because such a long-range (15$\sim$20~\AA) coupling itself is beyond any conventional magnetic 
models, it is not possible to obtain a significant conclusion from calculations based on the 
conventional magnetic theory.  

In Subsection~\ref{sec-g-c7} we demonstrated that the nonbonding state, $p_z^*$, is not induced by 
an interaction between the atoms in the majority sublattice but by an imbalance between the $p_z$ orbitals
of two sublattices due to the removal of one atom in the minority sublattice. 
We have shown via the TB model that, considering only the interaction between the first nearest 
neighbors (1NN), $p_z^*$ consists of the $p_z$-orbitals of only the atoms in the majority 
sublattice. Even when considering the interaction of the second nearest neighbors (2NN), the 
origin of the unpaired $p_z$-orbitals constrains the orbital components on the atoms of the 
minority sublattice as little as possible because any extra orbital components of $p_z^*$ on the 
minority sublattice would cause an extra imbalance between the two sublattices. Furthermore, as a 
point-defect state of a two-dimensional lattice resonating with energy of a perfect crystal, its 
wavefunction decays from the vacancy as $\sim$ $1/r$, keeping the orbital components on the minority 
sublattice as little as possible within a region of the decay length. According to these 
characteristics, we conclude that $p_z^*$ plays a key role not only in spin-polarization but also in 
magnetic ordering, based on the following analysis.  

Removing one atom from the minority sublattice causes one $p_z^*$ state and leaves one 
unpaired $p_z$ electron in the majority sublattice. Appearing near the Fermi level, the $p_z^*$ 
state will be filled by one remaining electron in spin polarization, leading to fractional MM 
distributed on the majority sublattice atoms. Removing two infinitely separated atoms in the 
minority sublattice could create two independent $p_z^*$ states and two free electrons. 

Imaging that the two vacancies are moved closer and closer until the two regions of each $p_z^*$ 
within the decay length overlap, the two independent $p_z^*$ states are therefore coherent. 
Within the overlap region, this favors keeping the wavefunction components as little as possible 
on the atoms of the minority sublattice because the $p_z^*$ states themselves do not originate 
from an interaction, but from an imbalance between the $p_z$-orbitals of the two 
sublattices, corresponding to a recombination of the unpaired $p_z$-orbitals of the majority 
sublattice atoms. The unpaired nature again constrains the sequence antisymmetric 
wavefunction extending from the overlap region to all the involved regions, leading to the two 
$p_z^*$ being coherent if they are within the decay length.  

As mentioned above, in addition to the $p_z^*$ in graphene with a vacancy, the antibonding state, 
$sp^{2*}$, which is short ranged, induces fractional magnetic moments on the three atoms 
surrounding the vacancy. Here $p_z^*$ can be seen as acting as an effective magnetic field to 
polarize the spin of the electrons filling on the $sp^{2*}$ state because both $p_z^*$ and 
$sp^{2*}$ originate all from the same vacancy. 

The above analysis, in two respects (spin-polarization and magnetic ordering), indicates a never 
before reported mechanism that is different from the conventional mechanism. Different models 
for magnetic ordering, such as the direct exchange model, the indirect exchange model, the 
superexchange model, the double exchange model and the RKKY model
can be all traced to the same mechanism with different coupling manners.
However, there are no localized MM in this new unusual magnetic 
mechanism, instead, there are fractional MM of the entire moment that are always aligned parallel
due to the antisymmetric wavefunctions. Note that the coherent origin of $p_z^*$ is not an 
interaction between vacancies but the imbalance between the $p_z$ orbitals of the two 
sublattices; as in the origin of an isolated $p_z^*$ state, recombining the unpaired $p_z$-orbitals 
caused by more vacancies keeps the wavefunction of $p_z^*$ as little as possible in the minority 
sublattice. 

The temperature plays a role in the magnetic ordering for this unusual mechanism: while it does 
not act as a factor to decouple the MM, but it does destroy the nonbonding state itself or 
does change the electron filling situation, that is, the involved state is shifted 
from being partially to fully filled or from being partially filled to completely empty. 
This is why such a weak magnetization 
can be bewilderingly observed at room temperature in proton-irradiated graphene. 

In this way, we can also explain the existence of a small window for vacancy concentration. A previous 
study concluded that, if vacancy concentration is too large, localized $sp$-electrons become 
delocalized and therefore destroy the FM ordering; therefore, there is a small window of vacancy 
concentration, up to which no ferromagnetism can be observed\cite{Fis15}. We have stressed that no really 
localized electrons exist in band calculations based on SEA. According to the above analysis, we 
propose an alternative possibility that large vacancy concentrations mean 
that more electrons are released and more $p_z^*$ states are formed. 
The $p_z^*$ states, saying two per unit cell, can 
make contact with each other to form two combined states with some energy splitting.
(if considering also 2NN interaction in TB model). 
If the energy splitting of the two combined states is larger than the 
spin splitting, the two electrons released by two vacancies will fully fill the lower combined state 
without spin polarization, therefore destroying the ferromagnetism. 

\section{Conclusions}
Based on first principle calculations, we traced for the first time 
the magnetism in nominally nonmagnetic materials 
to the antisymmetric orbital combination of the involved 
electronic states and proposed a never before reported magnetic mechanism 
distinctly different from the conventional theory. 
We emphasize that in these two respects, our conclusions are different from 
the previous investigations. 

We investigated the origin of magnetism in 
graphene with vacancies. It was shown that MM existed in all the investigated cases (for $2\times 
2$ $\sim$ 8$\times$8), varying between 1.0~$\mu_{\rm B}$ and 1.6~$\mu_{\rm B}$ and reaching a 
stable value of 1.3~$\mu_{\rm B}$ in the 7$\times$7-sized cell. The compelling evidence shows 
that the MM in materials with only $sp$ electrons can be traced to 
the antisymmetric manner of the wavefunctions of the involved electronic states,
the $sp^{2*}$-antibonding and $p_z^*$-nonbonding states which appear near the Fermi level. 
This is a conclusion distinctly different from the previous investigations as pointed out in
section Introduction. 
Removing 
one atom from graphene creates a vacancy, leaving three dangling bonds on three atoms pointing 
toward the vacancy and breaking the balance between the $p_z$ orbitals of the two 
sublattices. Consequently, the three $sp^{2}$ dangling bonds interact with each other to hybridize 
one bonding state and two antibonding states without JTD (for 2$\times$2 and 3$\times$3) or 
with JTD (for 4$\times$4 $\sim$ 8$\times$8); in addition, the unpaired $p_z$ orbital on the majority 
sublattice atoms forms one nonbonding state. Because the spatial wavefunctions of both the 
$sp^{2*}$-antibonding and $p_z^*$-nonbonding states are antisymmetric, their spin 
wavefunctions should be symmetric according to the electron exchange antisymmetric principle. 
Appearing near the Fermi level, $sp^{2*}$ and $p_z^*$ will be partially filled in spin 
polarization. This is the origin of $sp$-electron spin-polarization in proton-irradiated 
graphene. We emphasize that, unlike the conventional magnetic 
models, the $p_z^*$ induced MM are fractional MM of a whole moment 
distributed not on one atom but on all the involved atoms. 

The nonbonding state stems not from an interaction between atoms but from an imbalance 
between the $p_z$ orbitals of the two sublattices due to removing one atom from 
the minority sublattice. Therefore, in addition to contributing fractional MM, the nonbonding state plays a 
critical role in magnetic ordering. If the vacancy concentration is large enough to cause the 
vacancy-affected regions to overlap each other, the requirement of as little orbital components as 
possible on the minority sublattice in the overlap regions makes the vacancy-induced $p_z^*$ 
states coherent because more vacancies mean more unpaired $p_z$-orbitals, which require 
recombination. The coherent process in the overlap region therefore constrains the 
antisymmetric wavefunction covering all the vacancy-affected regions, consequently causing 
ferromagnetism according to the electron exchange antisymmetric principle. Therefore, we can understand 
how, in proton-irradiated graphene, such far-flung spins can be so strongly aligned 
and cannot be destroyed even by a high temperature ($>300$~K). Obviously, the connecting thread is 
a mechanism that is different from any previously published models for magnetic ordering.

This work was supported by NFSC (No.61274097) and NBRPC (No. 2015CB921401).

\end{document}